\documentclass[a4paper,11pt]{article}
\usepackage{pos}
\usepackage{wrapfig}

\title{
  Sparse modeling approach to extract spectral functions
  with covariance of Euclidean-time correlators of lattice QCD
}
\ShortTitle{
  Sparse modeling approach to extract spectral functions...
}

\author*[a]{Junichi Takahashi}
\author[b]{Hiroshi Ohno}
\author[c]{Akio Tomiya}

\affiliation[a]{Meteorological College, Japan Meteorological Agency,\\
7-4-81, Asahi-cho, Kashiwa, Chiba 277-0852, Japan}
\affiliation[b]{Center for Computational Sciences, University of Tsukuba,\\
1-1-1, Tennodai, Tsukuba, Ibaraki 305-8577, Japan}
\affiliation[c]{Faculty of Technology and Science, International Professional University of Technology,\\
3-3-1, Umeda, Kita-ku, Osaka, 530-0001, Japan}

\emailAdd{mhjkk-takahashi@met.kishou.go.jp}
\emailAdd{hohno@ccs.tsukuba.ac.jp}
\emailAdd{akio@yukawa.kyoto-u.ac.jp}

\abstract{
  We present our sparse modeling study to extract spectral functions from Euclidean-time correlation functions.
  In this study covariance between different Euclidean times of the correlation function is taken into account,
  which was not done in previous studies. In order to check applicability of the method,
  we firstly test it with mock data which imitate possible charmonium spectral functions.
  Then, we extract spectral functions from correlation functions obtained from lattice QCD at finite temperature.
}

\FullConference{The 40th International Symposium on Lattice Field Theory (Lattice 2023)\\
July 31st - August 4th, 2023\\
Fermi National Accelerator Laboratory\\}

\begin{document}
\maketitle

\section{Introduction}
Meson spectral functions play a crucial role to study properties of the hot and dense medium formed
in relativistic heavy ion collisions since they carry important theoretical information on probes of
the Quark-Gluon Plasma such as the thermal dilepton rate~\cite{McLerran:1984ay, Braaten1990.PhysRevLett.64.2242, Moore:2006qn}
and quarkonia~\cite{Matsui:1986dk}.
Especially, the low frequency part of the spectral function is associated
with transport coefficients (e.g., the heavy quark diffusion coefficient~\cite{Petreczky2006.PhysRevD.73.014508}),
which are important inputs for explaining transport phenomena in the experiments.
\\
\indent
Lattice QCD calculations, however, cannot obtain the spectral function directly but
it is accessible from the meson correlation
function $G$ of Euclidean time $\tau$ through the following relation:
\begin{equation}
  G(\tau)
  =\int d^{3}x
  \langle
  J_{H}(\tau,\vec{x})J_{H}(0,\vec{0})
  \rangle
  =\int^{\infty}_{0}d\omega K(\omega,\tau)\rho_{H}(\omega),
  \label{eq:G=intKrho}
\end{equation}
where $J_{H}$ represents the local meson operator of a channel $H$
and $K$ is the integration kernel defined by
\begin{equation}
  \displaystyle K(\omega,\tau)\equiv
  \frac{\cosh\left[
      \omega\left(
      \tau-\frac{1}{2T}
      \right)
      \right]}{\sinh\left(
    \frac{\omega}{2T}
    \right)}
  \label{eq:1_kernel}
\end{equation}
in the Euclidean time range $0\le\tau\le 1/T$ with temperature $T$.
\\
\indent
When the frequency $\omega$ is discretized, eq.~\eqref{eq:G=intKrho} can be simply written as a linear equation
\begin{equation}
  \vec{G}=K\vec{\rho},
  \label{eq:linear_eq}
\end{equation}
where $\vec{G}$ and $\vec{\rho}$ are $M$ and $N$ dimensional vectors, respectively, and
$K$ is an $M \times N$ matrix. For typical lattice QCD calculations the temporal lattice size, i.e.,
$M$ is of $O(10)$ while $N$ must be of $O(1000)$ for sufficiently good resolution of the spectral function.
Therefore, solving eq.~\eqref{eq:linear_eq} to extract the spectral function
is an ill-posed inverse problem.
\\
\indent
There are lots of previous studies on extracting spectral functions from lattice QCD data,
using various techniques based on different ideas
~\cite{ASAKAWA2001459,Ding2018.PhysRevD.97.094503,Brandt.PhysRevD.92.094510}.
Sparse modeling is one of such techniques, which was applied recently for the first time to lattice QCD data
to obtain spectral functions of the energy-momentum tensor and the shear viscosity~\cite{itou2020sparse}.
In this study we conduct a more comprehensive investigation into the applicability of sparse modeling.
Moreover, we also compare our results with those of one of the previous studies to properly estimate
the systematic uncertainty.

\section{Sparse modeling}
Extracting spectral functions by using sparse modeling has been proposed in condensed matter physics~\cite{Shinaoka.PhysRevB.96.035147,Otsuki.PhysRevE.95.061302}.
The following is a brief summary of the sparse modeling procedures in this study.
\begin{enumerate}
  \item Perform a singular value decomposition of the kernel $K$:
        \begin{equation}
          K=USV^{\mathrm{t}},
        \end{equation}
        where $S$ is a diagonal matrix composed of singular values,
        and $U$ and $V$ are $M\times M$ and $N\times N$ orthogonal matrices, respectively.
  \item Transform the basis of the correlation function $\vec{G}$ and the spectral function $\vec{\rho}$
        by $U^{\mathrm{t}}$ and $V^{\mathrm{t}}$, respectively:
        \begin{equation}
          \vec{G}^{\prime}\equiv U^{\mathrm{t}}\vec{G},
          \quad
          \vec{\rho}^{\prime}\equiv V^{\mathrm{t}}\vec{\rho}.
        \end{equation}
  \item Choose up to $L$-th largest singular values satisfied with the condition $s_{l}/s_{1}\ge 10^{-15}$,
        where $s_{l}$ is the $l$-th largest singular value, and
        drop the components of $\vec{\rho}^{\prime}$ and $\vec{G}^{\prime}$
        corresponding to the other small singular values, which reduces the size of $U$, $V$ and $S$
        to $M\times L$, $N\times L$ and $L\times L$, respectively.
  \item Construct the cost function $F(\vec{\rho}^{\prime})$
        from the square error and the L$_{1}$ regularization term:
        \begin{equation}
          F(\vec{\rho}^{\prime})
          =\frac{1}{2}(\vec{G}^{\prime}-S\vec{\rho}^{\prime})^{\mathrm{t}}
          U^{\mathrm{t}}C^{-1}U(\vec{G}^{\prime}-S\vec{\rho}^{\prime})
          +\lambda||\vec{\rho}^{\prime}||_{1}
          \equiv\chi^{2}(\vec{\rho}^{\prime})
          +\lambda||\vec{\rho}^{\prime}||_{1}.
          \label{eq:2_F(rho')}
        \end{equation}
        Here, $C$ in the first term is the covariance matrix defined by
        \begin{align}
          C_{ij}      & =\frac{1}{N_{\mathrm{conf}}(N_{\mathrm{conf}}-1)}
          \sum^{N_{\mathrm{conf}}}_{n=1}
          \left(G(\tau_i)-G^{(n)}(\tau_i)\right)\left(G(\tau_j)-G^{(n)}(\tau_j)\right), \\
          G(\tau_{i}) & =\frac{1}{N_{\mathrm{conf}}}
          \sum^{N_{\mathrm{conf}}}_{n=1}G^{(n)}(\tau_{i}),
        \end{align}
        where $N_{\mathrm{conf}}$ is the total number of gauge configurations
        and $G^{(n)}(\tau)$ is the value of the correlation function
        measured on the $n$-th gauge configuration.
        In the second term
        $||\cdot||_{1}$ stands for the L$_{1}$ norm defined by
        $||\vec{\rho}^{\prime}||_{1}\equiv\sum_{i=1}^{L}|\rho^{\prime}_{i}|$
        and $\lambda$ is a positive hyperparameter which controls the contribution of the L$_{1}$ regularization
        relative to the square error. Note that the commonly-used maximum entropy method has a different
        regularization term proportional to the Shannon-Jaynes entropy, which measures difference between
        an output spectral function and a default model which contains prior information.
  \item Estimate the optimal value of $\lambda$,
        $\lambda_\mathrm{opt}$,
        in the same way as the previous study~\cite{itou2020sparse}, i.e.,
        we vary $\chi^2(\vec{\rho}^{\prime})$ as a function of $\lambda$ and search for a kink (see fig.~\ref{fig:chi2-lambda_PoS}).
        \begin{figure}[b]
            \centering
            \includegraphics[width=0.47\linewidth]{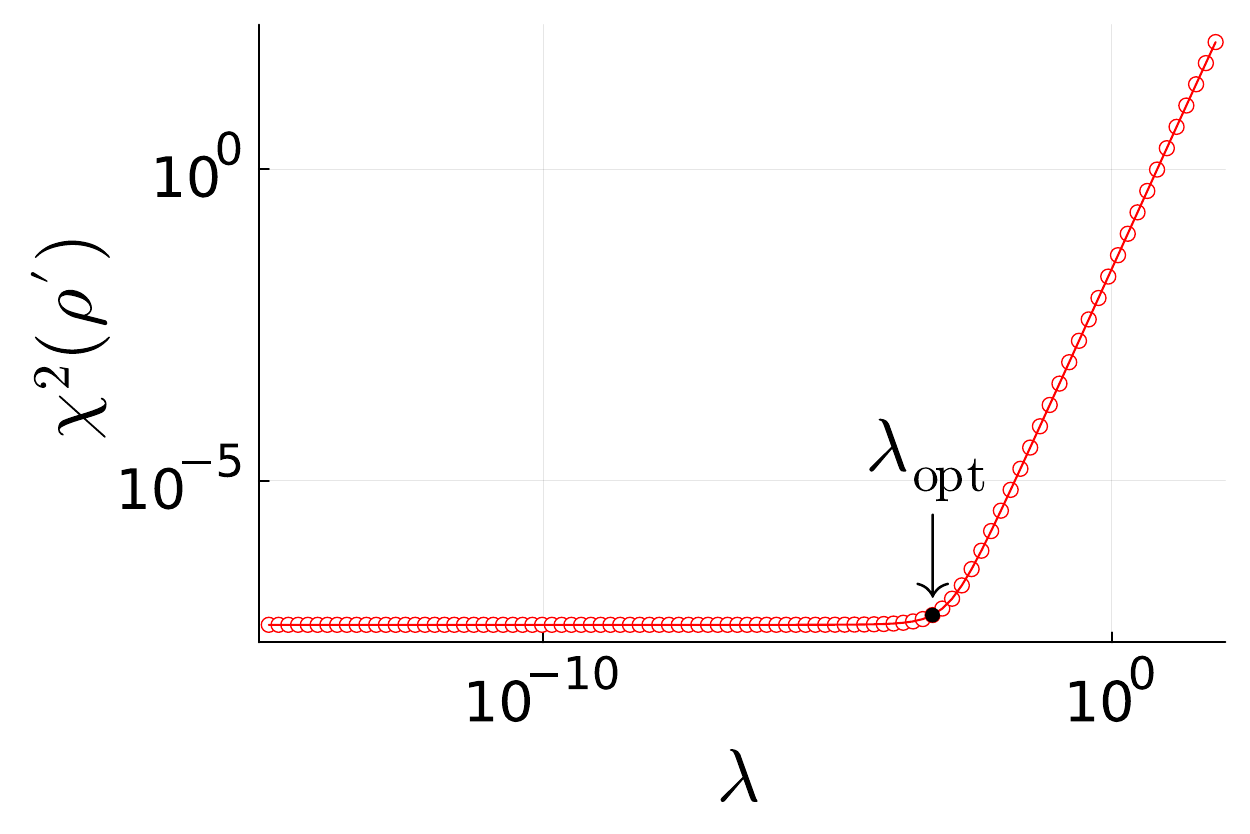}
            \caption{
            $\chi^2(\vec{\rho}^{\prime})$ in eq.~\eqref{eq:2_F(rho')} as a function of $\lambda$
            in the mock data test with $N_{\tau}=16$ mentioned later.
            The black filled circle represents the optimal value of $\lambda$,
            $\lambda_\mathrm{opt}$.
            }
            \label{fig:chi2-lambda_PoS}
        \end{figure}
  \item Find the most likely spectral function
        by minimizing the cost function $F(\vec{\rho}^{\prime})$
        using the ADMM algorithm~\cite{Boyd.MAL-016}
        with the positivity constraint $\rho_{i}\ge 0$.
\end{enumerate}
What differs from the previous study~\cite{itou2020sparse} is that the covariance matrix is considered
in $\chi^2(\vec{\rho}^{\prime})$ of the cost function so that correlation of $G$ between different
imaginary times is taken into account.

\section{Mock data tests}
\begin{figure}[tbp]
  \centering
  \begin{minipage}{.32\textwidth}
    \includegraphics[width=1.0\linewidth]{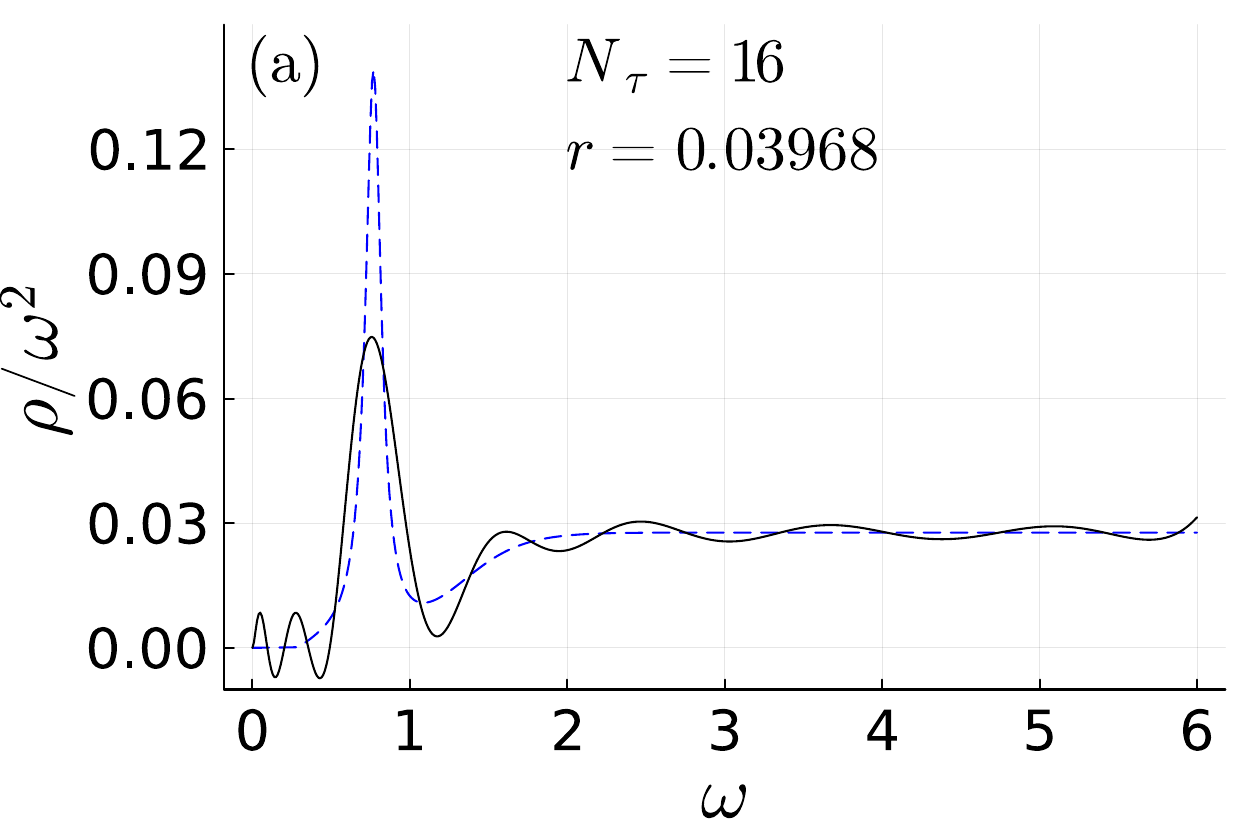}
  \end{minipage}
  \begin{minipage}{.32\textwidth}
    \includegraphics[width=1.0\linewidth]{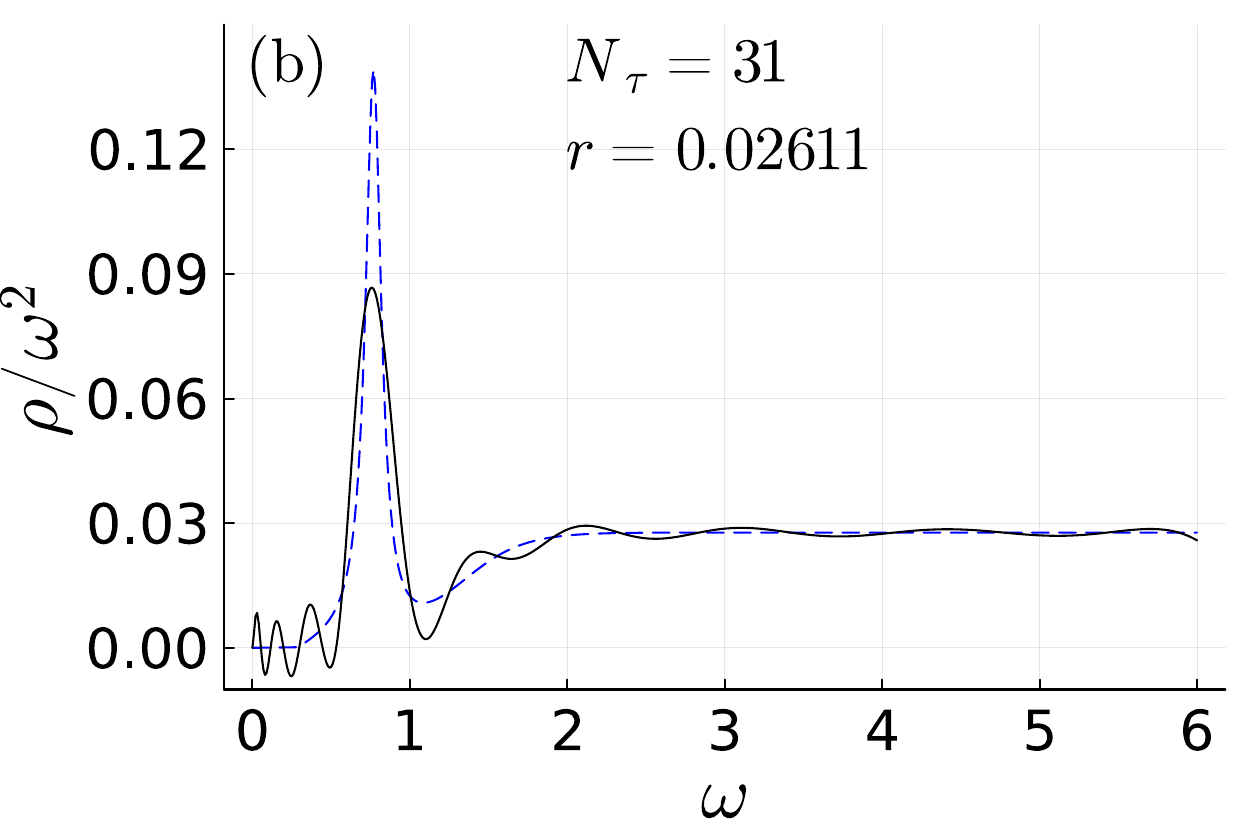}
  \end{minipage}
  \begin{minipage}{.32\textwidth}
    \includegraphics[width=1.0\linewidth]{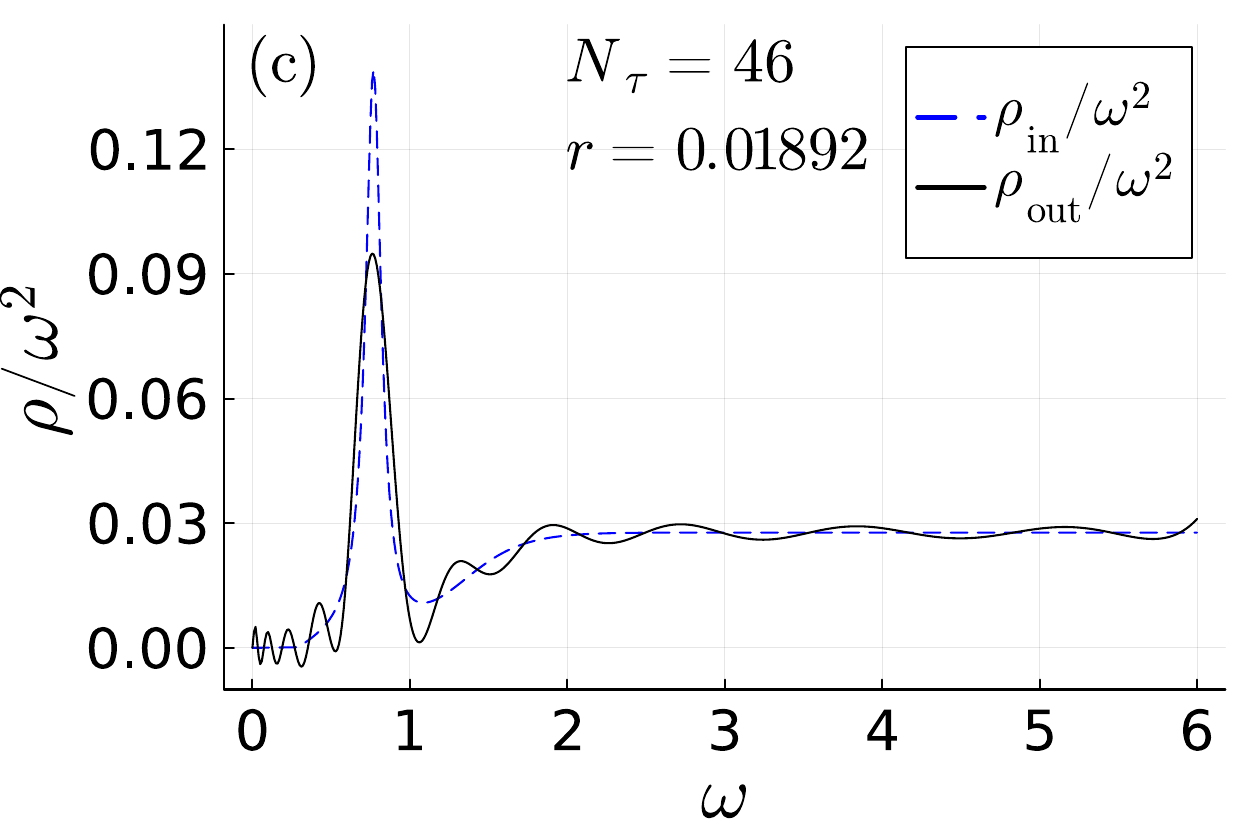}
  \end{minipage}
  \caption{
    Spectral functions calculated by using sparse modeling in the mock data tests
    with (a) $N_{\tau}=16$, (b) $N_{\tau}=31$ and (c) $N_{\tau}=46$.
    The blue dashed lines and red solid lines represent the input mock data $\rho_{\mathrm{in}}$
    and the output result $\rho_{\mathrm{out}}$.
  }
  \label{fig:spf_odn_PoS}
\end{figure}

\begin{figure}[tbp]
  \centering
  \begin{minipage}{.36\textwidth}
    \includegraphics[width=1.0\linewidth]{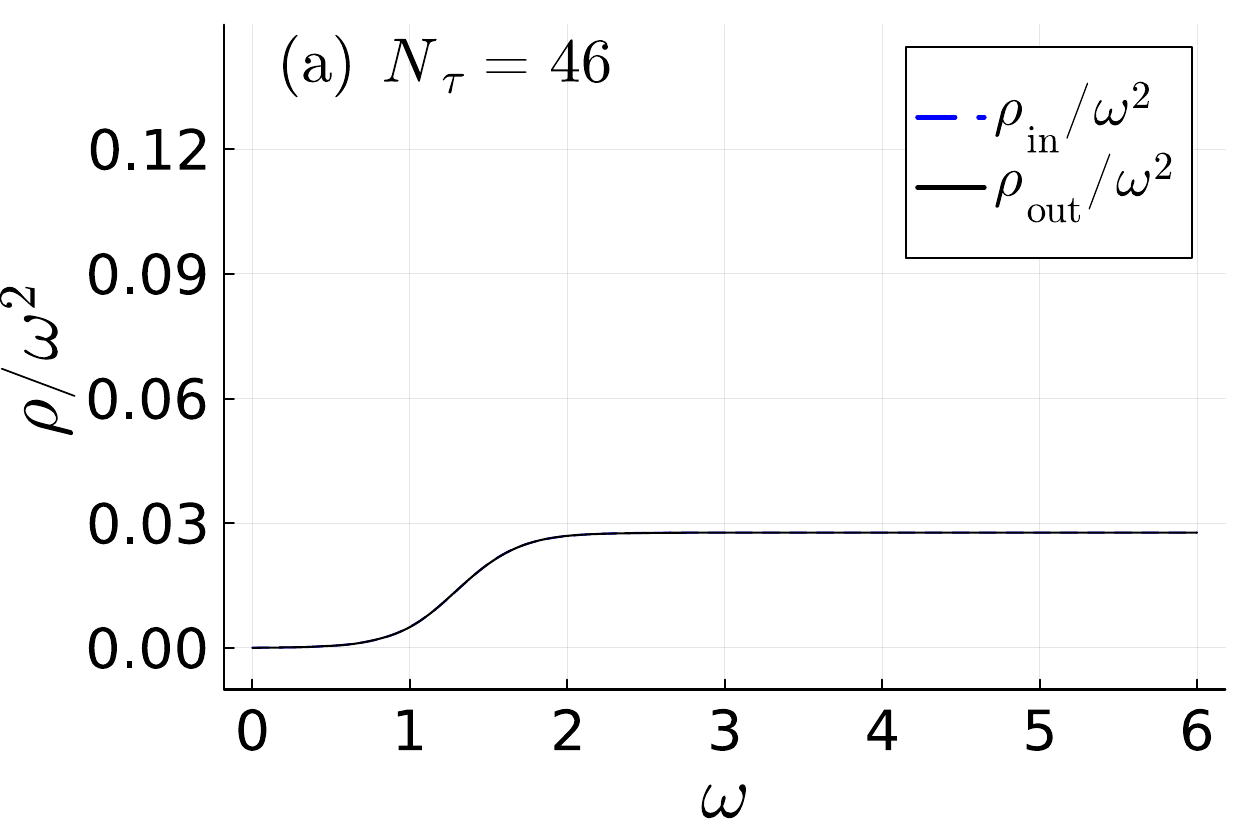}
  \end{minipage}
  \begin{minipage}{.36\textwidth}
    \includegraphics[width=1.0\linewidth]{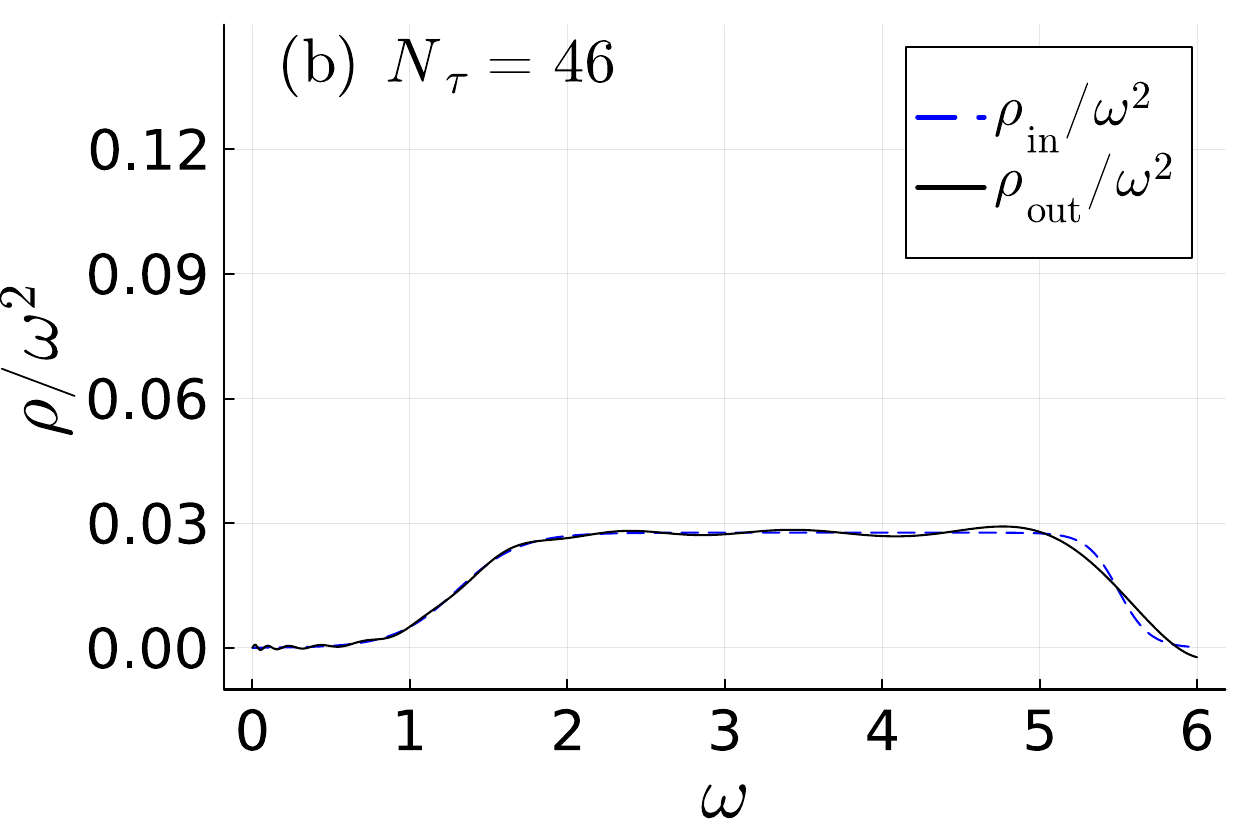}
  \end{minipage}
  \caption{
    The same as fig.~\ref{fig:spf_odn_PoS} but for the mock data with no peak.
    In figure (b), we further dumped $\rho_{\mathrm{in}}$ in the high-$\omega$ region.
    In figure (a), the two lines overlap each other almost perfectly.
  }
  \label{fig:spf_cont_c+d_Nt46_PoS}
\end{figure}

Before we analyze the actual lattice QCD data by using sparse modeling,
we test it with mock data which imitate possible charmonium spectral functions.
\\
\indent
We used the same mock data to those defined in ref.~\cite{Yamazaki.PhysRevD.65.014501}.
The input spectral function $\rho_{\mathrm{in}}(\omega)$ was set
to that in the vector channel of electron-positron pair annihilation, which is given by
\begin{equation}
  \rho_{\mathrm{in}}(\omega)
  =\frac{2\omega^{2}}{\pi}\left[
  F^{2}_{\rho}
  \frac{\Gamma_{\rho}(\omega)m_{\rho}}
  {(\omega^{2}-m_{\rho}^{2})^{2}+\Gamma_{\rho}^{2}(\omega)m_{\rho}^{2}}
  +\frac{1}{8\pi}\left(
  1+\frac{\alpha_{s}}{\pi}
  \right)\frac{1}{1+e^{(\omega_{0}-\omega)/\delta}}
  \right],
\end{equation}
where $F_{\rho}$ is the residue of $\rho$ meson resonance defined by
\begin{equation}
  \langle 0|\bar{d}\gamma_{\mu}u|\rho\rangle
  =\sqrt{2}F_{\rho}m_{\rho}\epsilon_{\mu}
  =\sqrt{2}f_{\rho}m_{\rho}^{2}\epsilon_{\mu},
\end{equation}
with the polarization vector $\epsilon_{\mu}$,
and the function $\Gamma_{\rho}(\omega)$ represents the threshold of decay from a $\rho$ meson to two $\pi$ mesons as
\begin{equation}
  \Gamma_{\rho}(\omega)
  =\frac{1}{48\pi}\frac{m_{\rho}^{3}}{F_{\rho}^{2}}\left(
  1-\frac{4m_{\pi}^{2}}{\omega^{2}}
  \right)^{3/2}\theta(\omega-2m_{\pi}),
\end{equation}
with the $\rho$ meson mass $m_{\rho}$ and the pion mass $m_{\pi}$.
We set the lattice spacing $a$ to 1 $\mathrm{GeV}^{-1}$.
The values of the parameter such as $m_{\rho}$ and $m_{\pi}$ are listed in table~\ref{tab:3_mock-data_param}.
\begin{table}[tpb]
  \centering
  \begin{tabular}{|c|c|c|c|c|c|}
    \hline
    $m_{\rho}$ & $m_{\pi}$ & $F_{\rho}$ & $\omega_{0}$ & $\delta$ & $\alpha_{s}$ \\
    \hline
    0.77       & 0.14      & 0.142      & 1.3          & 0.2      & 0.3          \\
    \hline
  \end{tabular}
  \caption{
    The values of parameters in $\rho_{\mathrm{in}}(\omega)$.
    The lattice spacing $a$ is set to 1 $\mathrm{GeV}^{-1}$.
  }
  \label{tab:3_mock-data_param}
\end{table}
\\
\indent
The central values of correlation function $G(\tau)$ were given
by integrating $\rho_{\mathrm{in}}K$, where $K=e^{-\omega\tau}$, over $\omega$.
Since this kernel is only an exponentially dumped function,
the imaginary time resolution $\Delta\tau$ was set
from $\tau_{\mathrm{max}}=\Delta\tau(N_{\tau}-1)$,
where $\tau_{\mathrm{max}}$ and $N_{\tau}$ represent
the maximum imaginary time length and the temporal lattice size, respectively.
In this study, we fixed $\Delta\tau$ to 0.5.
Errors of $G(\tau)$ were generated by gaussian random numbers
with the variance $\sigma(\tau)=b\cdot e^{a\tau}G(\tau)$
in order to incorporate the fact
that the error of lattice correlation functions increases as $\tau$ increases.
We used $a=0.1$ and $b=10^{-10}$, respectively.
In this test,
no correlation of $G(\tau)$ between different $\tau$ was considered,
i.e., $C$ is diagonal.
\\
\indent
In this study,
we consider that
the range of $\omega$ is from 0 to 6
and the number of points in $\omega$-space is $N_{\omega}=601$.
We performed tests on three different $N_{\tau}$
which were set to 16, 31 and 46.
In order to measure the difference
between the input data of the spectral function $\rho_{\mathrm{in}}$
and the corresponding output result $\rho_{\mathrm{out}}$,
the reconstruction error $r$ is defined by
$r=\sum^{N_{\omega}}_{j=1}\left((\rho_{\mathrm{in}}(\omega_{j})-\rho_{\mathrm{out}}(\omega_{j}))/\omega^{2}\right)^{2}$.
\\
\indent
Figures~\ref{fig:spf_odn_PoS}(a)-(c)
show the spectral function as a function of $\omega$ for $N_{\tau}=16$, 31 and 46,
respectively.
The blue dashed lines and red solid lines represent $\rho_{\mathrm{in}}$ and $\rho_{\mathrm{out}}$, respectively.
The values of the reconstruction error $r$ for each $N_{\tau}$ are shown in each figure.
The reconstruction error $r$ becomes smaller as $N_{\tau}$ becomes larger.
Unfortunately, even though the positivity condition is imposed in our analysis,
it is not satisfied in the low-$\omega$ region.
\\
\indent
We also tested the case where we intentionally removed the peak from $\rho_{\mathrm{in}}$.
Figures~\ref{fig:spf_cont_c+d_Nt46_PoS}(a) and (b)
show the resulting spectral functions as a function of $\omega$
with $N_\tau=46$.
In fig.~\ref{fig:spf_cont_c+d_Nt46_PoS}(b),
we further dumped $\rho_{\mathrm{in}}$ in the high-$\omega$ region.
In both cases,
the oscillations of $\rho_{\mathrm{out}}$ are weaker
than those in the case of $\rho_{\mathrm{in}}$ with a peak,
and the positivity condition is almost satisfied.

\section{Results from lattice data}
Next, we extracted the spectral function from actual lattice QCD data.

\begin{figure}[t]
    \centering
    \includegraphics[width=0.51\linewidth]{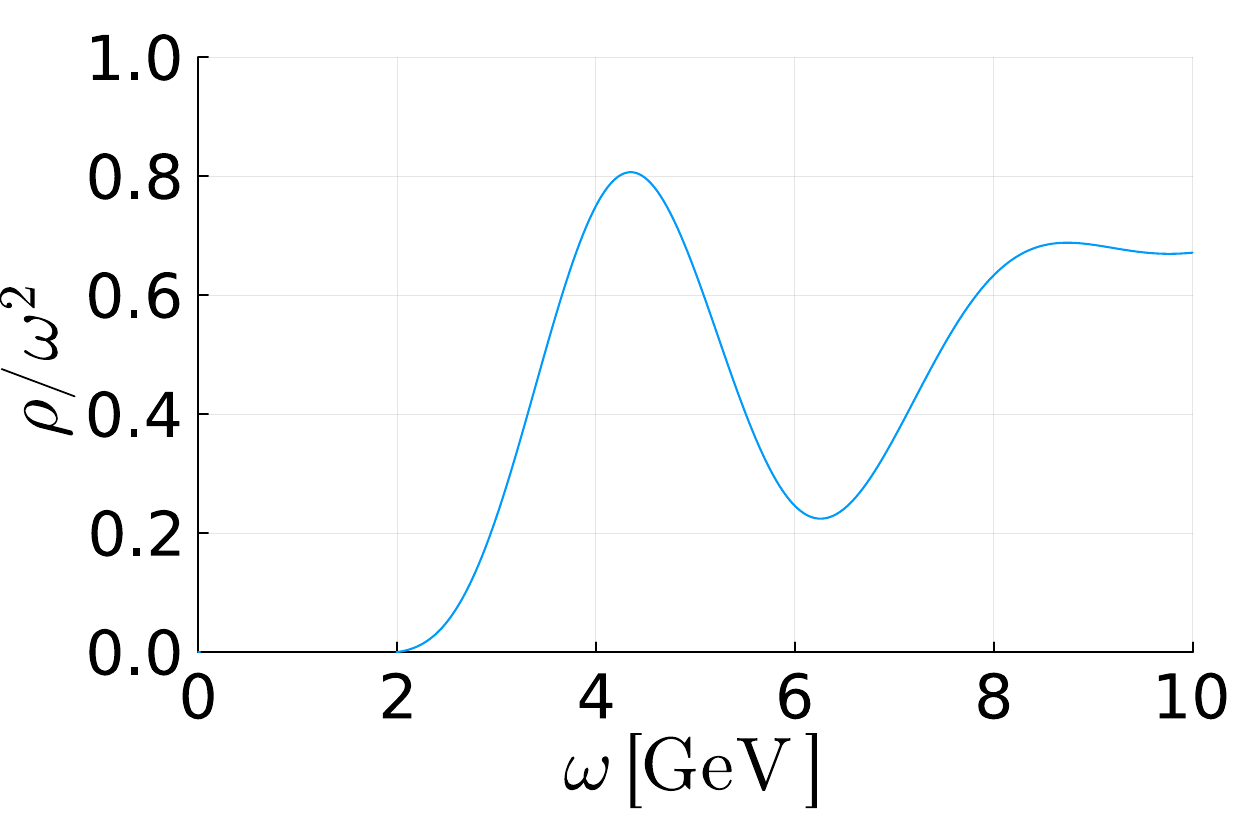}
    \caption{Spectral function obtained from the actual lattice QCD data.}
    \label{fig:spf_qLQCD_ve_T073_PoS}
\end{figure}

The lattice data used in this study were given in ref.~\cite{Ding2012.PhysRevD.86.014509},
where the correlation functions were measured with the $O(a)$-improved Wilson quark action
on quenched gauge configurations generated by using the standard plaquette gauge action.
The lattice spacing $a=0.010$ fm
and the corresponding $a^{-1}$ is about 18.97 GeV.
The spatial extent $N_{\sigma}$ and the temporal extent $N_{\tau}$
are 128 and 96, respectively.
This setup corresponds to temperature $T\simeq 0.73T_{\mathrm{c}}$.
We utilized meson correlation functions in the vector channel.
The number of gauge configurations is 234.
\\
\indent
The integration kernel is given in eq.~\eqref{eq:1_kernel},
which diverges at $\omega=0$.
Moreover,
the correlation function is influenced by lattice cutoff effects at small $\tau$ distances.
To address these issues,
we used a modified kernel and a modified spectral function defined by
\begin{align}
  \tilde{K}(\omega,\tau;\tau_{0})
  \equiv\displaystyle\omega^2\frac{K(\omega,\tau)}{K(\omega,\tau_{0})}
  =\omega^2\frac{\cosh\left[
      \omega\left(
      \tau-\frac{1}{2T}
      \right)
      \right]}{\cosh\left[
      \omega\left(
      \tau_{0}-\frac{1}{2T}
      \right)
      \right]},
  \quad
  \tilde{\rho}(\omega;\tau_{0})
  =\frac{\rho(\omega)}{\omega^2}K(\omega,\tau_{0}),
\end{align}
and we used the correlation function data from $\tau_{0}/a$ to $N_{\tau}/2$,
where $\tau_{0}/a$ was set to 4.
\\
\indent
Figure~\ref{fig:spf_qLQCD_ve_T073_PoS} shows our result of the spectral function.
The spectral function starts increasing around 2 GeV
and has a broad peak around 4 GeV.
The locations of the first peaks obtained from MEM is about 3.48 GeV
and the $J/\psi$ mass given by a single exponential fitting to the spatial correlation function
is about 3.47 GeV~\cite{Ding2012.PhysRevD.86.014509}.
Our result is a bit larger compared to these results.

\section{Summary and outlook}
We applied sparse modeling for extracting the spectral function
from the Euclidean-time meson correlation function.
Since the correlation function at different imaginary times correlate with each other,
in the cost function we introduced the square error term
with the covariance matrix so that covariance of the correlation function between
different Euclidean times were taken into account.
\\
\indent
First,
we tested sparse modeling with mock data of the spectral function
in the vector channel of electron-positron pair annihilation
and checked applicability of sparse modeling.
This test confirmed
that reconstruction error becomes smaller as the number of data points of
the correlation function becomes longer.
We also found that spectral functions with a peak violate the positivity condition,
while those with no peak are almost positive.
Then,
we tried to extract the spectral function from the vector charmonium correlation function
obtained from lattice QCD.
Then, We got a spectral function with a broad peak around 4 GeV,
which is a bit larger compared to the results in the previous study.
\\
\indent
Base on this study,
it is necessary to further investigate how to keep positivity of spectral functions,
how to find the optimal $\lambda$ and convergence of ADMM iterations.
Investigating whether transport peaks appear in the spectral functions
at higher temperature is also our future work.

\acknowledgments
We deeply grateful to H.-T. Ding for sharing lattice data.
The work of A.T. was partially supported by JSPS KAKENHI Grant Numbers 20K14479, 22H05111 and 22K03539.
A.T. and H.O. were partially supported by JSPS KAKENHI Grant Number 22H05112.
This work was partially supported by MEXT as ``Program for Promoting Researches on the Supercomputer Fugaku'' (Grant Number JPMXP1020230411, JPMXP1020230409).

\bibliographystyle{JHEP}
\bibliography{453_SpM}

\end{document}